\documentstyle[preprint,aps]{revtex}

\begin{document}
\tighten
\draft
\preprint{
 \parbox[t]{50mm}{hep-ph/9606214 % archive
 \parbox[t]{50mm}{DPNU-96-21\\
}
}}

\title{The Influence of the Full Vertex and Vacuum Polarization 
       on the Fermion Propagator in QED3}

\author{P. Maris\cite{emmar}}
\address{Department of Physics, Nagoya University, 
Nagoya 464-01, Japan}
\date{April 1996}
\maketitle
\begin{abstract}
We investigate the influence of the full vacuum polarization and
vertex function on the fermion propagator, using the coupled
Dyson--Schwinger equations for the photon and fermion propagator. We
show that, within a range of vertex functions, the general behavior of
the fermion propagator does not depend on the exact details of the
vertex, both in the massless and in the massive phase. Independent of
the precise vertex function, there is a critical number of fermion
flavors for dynamical mass generation in $(2+1)$-dimensional QED. A
consistent treatment of the vacuum polarization is essential for these
results.
\end{abstract}
%

%%%%%%%%%%%%%%%%%%%%%%%%%%%%%%%%%%%%%%%%%%%%%%%%%%%%%%%%%%%%%%%%%%%%%%%%%%%%%
\section{Introduction}

Quantum electrodynamics in 2 space- and 1 time-dimension, QED3, has
several interesting features. It exhibits dynamical mass generation
\cite{Pi84,Appetal,Na89,HoMa89,DKK89,Penetal88,PW91CPW92,AJM90,KoNa92,BuRo91} 
and confinement \cite{BuPrRo92,GrSeSo95,Ma95}, similar to QCD. 
Furthermore, it is superrenormalizable, so it does not suffer from the
ultraviolet divergences which are present in QED4. The coupling
constant is dimensionful, and provides us with a mass scale, even if
we consider massless fermions. This energy scale plays the role of the
QCD scale $\Lambda_{QCD}$, in the sense that it sets the scale for
confinement and dynamical mass generation. Thus it is a very
interesting model to study these nonperturbative phenomena.
 
QED3 also has some applications in condensed matter physics, where it
can be regarded as an effective theory for more realistic microscopic
models \cite{DoMa,AbCa95}. Especially, since the discovery of
high-$T_c$ superconductivity and the fractional quantum Hall effect,
these kinds of models have attracted more attention.

In this paper, we consider QED3 with $N$ fermion flavors of
four-component spinors. Such a model of QED3 is chirally symmetric in
the absence of a bare fermion mass term, $m_0 \bar \psi \psi$, in
contrast to the (2+1)-dimensional gauge theory with two-component
fermions, where we cannot define chiral symmetry
\cite{Pi84,Appetal,ABKW86}. Similar to the four-dimensional case
\cite{Miretal}, the chiral symmetry of QED3 may be broken
spontaneously due to the dynamical generation of a fermion mass. The
question of whether or not chiral symmetry is broken for all values of
$N$, the number of fermion flavors, is very interesting.

We address this question by analyzing the behavior of the full fermion
propagator nonperturbatively, using its Dyson--Schwinger equation.
However, this equation cannot be solved without truncating the
infinite set of Dyson--Schwinger equations, since it involves the full
photon propagator and the full vertex function. Different ways of
truncating this equation give rise to different results concerning the
question of whether there is a critical number of fermion flavors for
dynamical mass generation
\cite{Appetal,Na89,Penetal88,PW91CPW92,AJM90,KoNa92}. In this paper we
try to resolve the controversy.

Using the bare vertex approximation and the one-loop vacuum
polarization, Appelquist {\em et al.} \cite{Appetal} have shown that
there is a finite critical number of flavors $N_c$, above which the
chiral symmetry is restored. They found in Landau gauge a critical
value of $N_c = 32/\pi^2 \simeq 3.24$. This approach is based on a
$1/N$ expansion, and including the next-to-leading-order terms
\cite{Na89}, it was found that the critical number changes to
$N_c=128/3\pi^2 \simeq 4.32$.

However, such a simple treatment was criticized by Pennington {\em et
al.} \cite{Penetal88}, since the effects of the wavefunction
renormalization were not taken into account. The problem is that
formally the wavefunction renormalization is of order $1 + {\cal
O}(1/N)$, but in the infrared region, this wavefunction
renormalization tends to vanish. Because of this non-uniformity in the
$1/N$ expansion, an approximation based on an expansion in $1/N$ might
not be very reliable. Taking into account wavefunction
renormalization, Pennington {\em et al.} found chiral symmetry
breaking for all numbers of fermion flavors. However, as we show,
their approach also has some inconsistency.

There have been several attempts to resolve the problem, by means of
more sophisticated Ans\"atze for the full vertex
\cite{PW91CPW92,AJM90,KoNa92,BuRo91}, and by use of a so-called nonlocal
gauge-function \cite{Si90,KM94,GuHaRe95}, but none of them are
completely satisfactorily. Also other methods, such as the inversion
method \cite{Ko95}, $\epsilon$-expansion \cite{epsilon}, and lattice
calculations \cite{DKK89}, have not given a final answer to this
question, although the lattice results favor the existence of a finite
critical number of fermion flavors \cite{DKK89,KiKi96}. The difficulty
is that it is numerically very difficult to observe the exponential
decrease of the dynamical mass for increasing $N$ found in
\cite{Penetal88,PW91CPW92}.

In our paper, the existence of a finite critical number of flavors is
confirmed by analyzing the {\em coupled} Dyson--Schwinger equations,
for both the fermion and the photon propagator, using several
different approximations for the full vertex. We show that, within a
certain class of vertex Ans\"atze, there is a critical number of
fermion flavors for dynamical mass generation in QED3. This number
does not depend very strongly on the precise form of the vertex
Ansatz. Also the general behavior of the fermion propagator does not
depend on the exact details of the Ansatz. The essential point is to
take into account the full vacuum polarization in a consistent way.

In the next section, we describe the formalism in more detail,
together with our truncation scheme. Before addressing the question of
chiral symmetry breaking, we first analyze the full fermion propagator
in the massless phase, see Sec.~\ref{secsymphase}. This already yields
some nontrivial results, namely that the structure of the massless
full fermion propagator is almost independent of our Ansatz. In
Sec.~\ref{secbroken}, we discuss dynamical mass generation, and show
that there is a critical number of fermion flavors for chiral symmetry
breaking, $N_c \simeq 3.3$, almost independent of our choice for the
vertex. Finally we give some concluding remarks in Sec.~\ref{secconc}.

%%%%%%%%%%%%%%%%%%%%%%%%%%%%%%%%%%%%%%%%%%%%%%%%%%%%%%%%%%%%%%%%%%%%%%%%%%%%%
\section{Formalism}

\subsection{Three-dimensional QED with $N$ massless fermion flavors}

We consider QED3 with $N$ massless fermion flavors, and choose to work
in Euclidean space, ignoring the issues discussed in \cite{Ma95}. The
Lagrangian in a general covariant gauge is given by
\begin{eqnarray}
{\cal L} &=& \sum_{i=0}^{N} 
        \bar\psi_i( i \not{\!\partial} + e\not{\!\!A})\psi_i 
        + {\textstyle{1\over4}} F_{\mu\nu}^2 
        + {\textstyle{1\over{2a}}} (\partial_\mu A_\mu)^2 \,.
\end{eqnarray}
We use four-component spinors for the fermions, and accordingly a
four-dimensional representation for the $\gamma$-matrices. With such a
representation we can define chirality just as in four-dimensional
QED. This chiral symmetry can be broken dynamically by generation of a
mass for the fermions. In this formulation, there can also be a parity
breaking mass term, which conserves the chiral symmetry, but it is
known that such a mass is not generated dynamically \cite{ABKW86}.

We have $N$ fermion flavors, and consider both the large $N$ limit, as
well as the quenched limit, $N \downarrow 0$. In the quenched limit,
the mass scale is defined by the dimensionful coupling constant $e^2$,
and there are no free parameters. Outside the quenched limit, $N$, the
number of fermion flavors is the only free parameter. When using the
large $N$ expansion \cite{Appetal}, we keep the product $N e^2$
finite, and it is most convenient to define the mass scale by
\begin{eqnarray}
  \alpha &=& \frac{N e^2}{8} \,.
\end{eqnarray}

\subsection{Dyson--Schwinger equations for the propagators}

The Dyson--Schwinger (DS) equation for the fermion propagator is given
by
\begin{eqnarray}      
 S^{-1}(p) &=& S^{-1}_0(p) - 
        e^2 \int\!\frac{{\rm d^3}k}{(2\pi)^3} 
        \gamma_\mu S(k) \Gamma_\nu(p,k) D_{\mu\nu}(p-k)\,.
\end{eqnarray}
Since parity is not broken dynamically, we can decompose the fermion
propagator into
\begin{eqnarray}      
 S^{-1}(p) &=& A(p) {\,\not\!p} + B(p) \,,
\end{eqnarray}
and rewrite the DS equation into
\begin{eqnarray}      
 A(p) &=& 1 + 
 \frac{e^2}{p^2} \int\!\frac{{\rm d^3}k}{(2\pi)^3}{\textstyle\frac14} 
 {\rm Tr}[{\,\not\!p}\,\gamma_\mu S(k)\Gamma_\nu(p,k) D_{\mu\nu}(p-k)]\,,
\\
 B(p) &=& e^2 \int\!\frac{{\rm d^3}k}{(2\pi)^3} 
 {\textstyle\frac14} 
 {\rm Tr}[\gamma_\mu S(k)\Gamma_\nu(k,p)D_{\mu\nu}(p-k)]\,.
\end{eqnarray}
The problem in analyzing this equation is the full vertex and full
photon propagator. For the photon propagator we also have a DS
equation, namely
\begin{eqnarray}      
 D^{-1}_{\mu\nu}(q) &=& {D^{-1}_0}_{\mu\nu}(q) - 
        e^2 \int\!\frac{{\rm d^3}k}{(2\pi)^3} 
        \gamma_\mu S(k) \Gamma_\nu(k,p-k) S(p-k)\,,
\end{eqnarray}
without introducing new unknown functions. It is more convenient to
write this last expression in terms of the vacuum polarization tensor
$\Pi_{\mu\nu}(q)$
\begin{eqnarray}      
 \Pi_{\mu\nu}(q) &=& e^2 \int\!\frac{{\rm d^3}k}{(2\pi)^3} 
        \gamma_\mu S(k) \Gamma_\nu(k,p-k) S(p-k)\,.
\end{eqnarray}
Since the longitudinal part of the photon propagator is not affected
by the interactions, because of gauge invariance, we can write the
full photon propagator in a general covariant gauge 
\begin{eqnarray}      
 D_{\mu\nu}(q) &=& 
  - \left( \delta_{\mu\nu} - \frac{q_\mu q_\nu}{q^2}\right)
    \frac{1}{q^2 + \Pi(q)} - a \frac{q_\mu q_\nu}{q^4} \,,
\end{eqnarray}
with the vacuum polarization $\Pi(q)$ defined by
\begin{eqnarray}      
 \Pi_{\mu\nu}(q) &=& 
  \left( \delta_{\mu\nu} - \frac{q_\mu q_\nu}{q^2}\right) 
  \Pi(q) \,.
\end{eqnarray}
The vacuum polarization tensor has an ultraviolet divergence, which
can be removed by a gauge-invariant regularization scheme. However,
this divergence is only present in the longitudinal part, so by
contracting $\Pi_{\mu\nu}(q)$ with
\begin{eqnarray}      
   q^2 \delta_{\mu\nu} - 3 \frac{q_\mu q_\nu}{q^2} \,,
\end{eqnarray}
we can project out the finite vacuum polarization $\Pi(q)$
\cite{BuPrRo92}. So, the coupled DS equations for the photon and
fermion propagator form a set of three coupled equations for three
scalar functions, and the only unknown function is the full vertex
function. Note that both the DS equation for the fermion propagator,
and the one for the photon propagator, are exact.

In principle, we could write down a DS equation for the full vertex
function as well, but this will not lead to a closed set of equations:
the DS equation for the vertex involves a four-point function, and so
on. The full set of DS equations forms an infinite hierarchy of
coupled integral equations for the Green's functions. In order to
solve the DS equation for a particular Green's function, we have to
truncate or approximate this infinite set of equations. For
calculating the propagators, we must find a reasonable approximation
for the full vertex function $\Gamma^\mu(p,k)$.

\subsection{Truncation scheme}

Now, what is a reasonable approximation for the full vertex function? 
The most simple and in some sense ``natural'' approximation is to take
the leading order perturbative vertex
\begin{eqnarray}
 \Gamma_\mu &\rightarrow& \gamma_\mu \,.
\end{eqnarray}
This truncation is commonly used in studies of the fermion DS
equation, and it is usually referred to as ladder or rainbow
approximation, since it generates rainbow diagrams in the fermion DS
equation, and ladder diagrams in the Bethe-Salpeter equation for the
fermion--anti-fermion bound state amplitude. In principle there is a
systematic way to improve this truncation, namely by taking the
next-to-leading order vertex function, and so on.

The obvious disadvantage of this approach is that it relies on
perturbation theory, whereas the DS equations are nonperturbative
equations. Until one performs the complete next-to-leading order
calculation, one does not know how reliable the leading order
calculation is. Even after obtaining such a next-to-leading order,
there remains some doubt about the validity of this approach, due to
the non-uniformity of the $1/N$ expansion.

Furthermore, as a consequence on the one hand of using a
nonperturbative method to calculate some Green's functions, and on the
other hand of employing perturbation theory for other Green's
functions, one violates the Ward--Takahashi (WT) identities relating
these Green's functions, and loses gauge covariance. To be specific,
the WT identity relating the full vertex to the fermion propagator
\begin{eqnarray}
 q_{\mu}\Gamma_\mu &=& S^{-1}(p) - S^{-1}(k) \,,
\end{eqnarray}
is an exact identity, which holds order-by-order in perturbation
theory, but if one uses perturbation theory to approximate the full
vertex, and the truncated DS equation to calculate the fermion
propagator, one will (in general) violate this identity.

Instead of using a perturbative approximation for the vertex function,
one can also use other nonperturbative information to make an Ansatz
for the full vertex. The full vertex can be decomposed into 12
different Lorentz structures, and four of them are uniquely determined
by the WT identity \cite{BaCh80}. So by imposing the WT identity, we
can write the full vertex as
\begin{eqnarray} \label{bachver}
 \Gamma_\mu &=& 
   {\textstyle{\frac{1}{2}}}(A(p)+A(k))\gamma_\mu 
   + {\textstyle{\frac{1}{2}}}({\not\! p} + {\not\! k})
     (p_\mu+k_\mu)\frac{A(p)-A(k)}{p^2-k^2} 
\nonumber \\ &&
   - (p_\mu+k_\mu)\frac{B(p)-B(k)}{p^2-k^2} + \ldots \;,
\end{eqnarray}
where the dots represent the part of the vertex not constrained by the
WT identity. Now one can simply neglect that unconstrained part, and
take the above expression as Ansatz (usually called Ball--Chiu vertex)
for the full vertex. However, a perturbative calculation shows that
the unconstrained part is not zero \cite{KiRePe95}, and by just
neglecting it you never see what you are throwing away.

The WT identity is not the only requirement one can impose on the full
vertex function. Other requirements are that it reduces to the bare
vertex in the weak coupling limit, and multiplicative
renormalizability also restricts the full vertex, even in a
superrenormalizable theory like QED3. Furthermore, any Ansatz should
have the correct symmetry properties. However, all these constraints
do not uniquely determine the full vertex, they all leave parts of the
vertex undetermined. So the results might depend heavily on the
particular choice for the vertex. Another disadvantage of using a
vertex Ansatz is that it is not possible to improve the approximation
in a systematic way.

\subsection{Vertex Ans\"atze}

In this paper, we use several different Ans\"atze for the full vertex,
including a bare vertex, and compare the results to see how much
influence these different choices have on the propagators. For the
sake of simplicity, we consider the form
\begin{eqnarray}
\label{genansatz}
 \Gamma_\mu(p,k) &=& f(A(p),A(k),A(p-k)) \gamma_\mu \,,
\end{eqnarray}
with the restriction that for $A(p) \equiv 1$, the full vertex reduces
to the bare vertex. This automatically ensures that in the weak
coupling limit, the full vertex reduces to the bare one. Such an
Ansatz has the same tensor structure as the bare vertex, and by
restricting ourselves to this tensor structure, we simplify the
numerical integrations considerably. It is also generally expected
that this tensor structure plays the most dominant role in the full
vertex.
 
To be more specific, we will analyze the coupled DS equations for the
propagators, using the following choices for $f(A(p),A(k),A(p-k))$:
\begin{enumerate}
\item the bare vertex: $ f(A(p),A(k),A(q)) = 1$ 
\label{bare}
\item a simple Ansatz inspired by the Ball--Chiu vertex 
      Eq.~(\ref{bachver}) \cite{BaCh80}: 
      $f = \frac{1}{2}(A(p)+A(k))$
\label{simple}
\item $f = A(p)A(k)/A(q)$
\label{apkq}
\item $f = \frac{1}{4}(A(p)+A(k))^2$
\label{quad}
\item $f = A(p)A(k)$.
\label{prod}
\end{enumerate}
Of these choices, only the first two are actually motivated on
physical arguments. The last one is primarily motivated by the fact
that the resulting truncated DS equations are easy to solve
analytically (at least in the symmetric phase). The choices \ref{apkq}
and \ref{quad} are merely added in order to see how much the results
depend on our particular choice; Ansatz \ref{apkq} is based on
\ref{prod}, but linear in the wavefunction renormalization, as it is
supposed to be, whereas Ansatz \ref{quad} is the quadratic form of
Ansatz \ref{simple}. The last two Ans\"atze are also inspired by the
suggestion in \cite{AJM90} that the effective vertex correction in the
fermion DS equation is quadratic in the wavefunction renormalization.

Finally, we have to choose a gauge, and since the Landau gauge is the
most convenient and commonly used gauge, we use it. Using the Ansatz
for the vertex we have just described, the DS equations for the
propagators reduce to
\begin{eqnarray}
\label{DSeqA}  
 A(p) &=& 1 + \frac{2\,e^2}{p^2} \int\!\frac{{\rm d^3}k}{(2\pi)^3}  
        \frac{A(k) (p\cdot q)(k\cdot q)}{A^2(k)k^2 + B(k)} 
        \frac{f(A(p),A(k),A(q))}{q^2(q^2+\Pi(q))} \,, \\
\label{DSeqB} 
 B(p) &=& 2\,e^2 \int\!\frac{{\rm d^3}k}{(2\pi)^3} 
        \frac{B(k)}{A^2(k)k^2 + B(k)} 
        \frac{f(A(p),A(k),A(q))}{q^2+\Pi(q)} \,, \\
\label{DSeqPi} 
 \Pi(q) &=& N e^2 \!\int\!\frac{{\rm d^3}k}{(2\pi)^3} 
   \left(2k^2 - 4 k\cdot q - \frac{6k\cdot q}{q^2}\right) 
   \frac{A(k)}{A^2(k)k^2 + B^2(k)} \,
   \frac{A(p)f(A(p),A(k),A(q))}{A^2(p)p^2 + B^2(p)}  ,
\end{eqnarray}
with $q=p-k$.

%%%%%%%%%%%%%%%%%%%%%%%%%%%%%%%%%%%%%%%%%%%%%%%%%%%%%%%%%%%%%%%%%%%%%%%%%%%%%
\section{Symmetric Phase}
\label{secsymphase}

First, we consider the massless fermion phase, so $B(p)\equiv 0$,
which is always a solution of Eq.~(\ref{DSeqB}). This reduces the
problem to solving two coupled equations, one for the wavefunction
renormalization $A(p)$, and one for the vacuum polarization $\Pi(p)$.

\subsection{Analytical results}

With $B(p)\equiv 0$, the vacuum polarization reduces to
\begin{eqnarray}      
 \Pi(q) &=& N e^2 \int\!\frac{{\rm d^3}k}{(2\pi)^3} 
        \frac{2k^2 - 4 k\cdot q - 6(k\cdot q)/q^2}
             {k^2 (k+q)^2}
        \frac{f(A(k+q),A(k),A(q))}{A(k)A(k+q)}  \,.
\end{eqnarray}
We first consider Ans\"atze \ref{apkq} and \ref{prod}, because
the vacuum polarization can then be calculated analytically.
With Ansatz \ref{apkq}, the vacuum polarization becomes
\begin{eqnarray}      
 \Pi(q) &=& N e^2 q / (8 A(q))  = \alpha q / A(q)  \,,
\end{eqnarray}
whereas Ansatz \ref{prod} leads to
\begin{eqnarray}      
 \Pi(q) &=& N e^2 q / 8  = \alpha q \,.
\end{eqnarray}
In the massless phase, the equation for the wavefunction
renormalization becomes
\begin{eqnarray}
  A(p) &=& 1 + \frac{2\,e^2}{p^2} \int \frac{{\rm d}^3k}{(2\pi)^3}
   \frac{f(A(p),A(k),A(q))}{A(k) k^2}
   \frac{(p\cdot q)(k\cdot q)}{q^2(q^2+\Pi(q))} \,.
\end{eqnarray}
For the Ans\"atze \ref{apkq} and \ref{prod}, this can now be reduced
to
\begin{eqnarray}  \label{apkqsymeq}
  A(p) &=& 1 + A(p) \frac{16\alpha}{N p^2} \int \frac{{\rm d}^3k}{(2\pi)^3}
   \frac{1}{A(q) q^2+ \alpha q} 
   \frac{(p\cdot q)(k\cdot q)}{k^2 q^2}\,,
\\
\label{prodsymeq}
  A(p) &=& 1 + A(p) \frac{16\alpha}{N p^2} \int \frac{{\rm d}^3k}{(2\pi)^3}
   \frac{1}{q^2+\alpha q} 
   \frac{(p\cdot q)(k\cdot q)}{k^2 q^2} \,,
\end{eqnarray}
where we use $\alpha = N \, e^2 /8$ rather than $e^2$ to define the
energy scale. This last equation can be solved exactly
\begin{eqnarray}  \label{apkqsymsol}
  A^{-1}(p) &=& 1 - \frac{2\alpha}{N p^2\pi^2} 
   \int_0^\infty \frac{{\rm d}q}{q^2+\alpha q} 
   \left(q^2 - p^2 - \frac{q^4 - p^4}{2pq} \ln{\frac{p+q}{|p-q|}}\right) \,,
\end{eqnarray}
whereas Eq.~(\ref{apkqsymeq}) gives a nonlinear integral equation for
$A(p)$
\begin{eqnarray}  \label{prodsymsol}
  A^{-1}(p) &=& 1 - \frac{2\alpha}{N p^2\pi^2} 
   \int_0^\infty \frac{{\rm d}q}{A(q) q^2+\alpha q} 
   \left(q^2 - p^2 - \frac{q^4 - p^4}{2pq} \ln{\frac{p+q}{|p-q|}}\right) \,,
\end{eqnarray}
and thus gives an implicit relation for the wavefunction
renormalization.

However, both Ans\"atze lead to the same explicit expression, if we
make one further approximation, which is commonly used in this context.
In the infrared region, $q \ll \alpha$, the vacuum polarization
dominates: the denominators, $A(q) q^2 + \alpha q$ in
Eq.~(\ref{apkqsymeq}), and $q^2 + \alpha q$ in Eq.~(\ref{prodsymeq}),
both behave like $\alpha q$, whereas for large momenta, $q \gg
\alpha$, the wavefunction renormalization is almost equal to one, 
and the contribution to these denominators coming from the vacuum
polarization can be neglected.  So in both cases the wavefunction
renormalization behaves like
\begin{eqnarray} 
  A^{-1}(p) &\simeq& 1 - \frac{2\alpha}{N p^2\pi^2} 
   \left( \int_0^\alpha \frac{{\rm d}q}{\alpha q} 
        + \int_\alpha^\infty \frac{{\rm d}q}{q^2} \right)
   \left(q^2 - p^2 - \frac{q^4 - p^4}{2pq} \ln{\frac{p+q}{|p-q|}}\right)
\nonumber \\ 
\label{analres}
&=& 1 - \frac{1}{N \pi^2} 
 \left( 1 -\frac{\alpha^2}{3p^2} 
        - \frac{3p^4 + 6p^2\alpha^2 - \alpha^4}{6p^3\alpha}
        \ln{\frac{\alpha+p}{|\alpha-p|}}
        - \frac{4}{3} \ln{\frac{|\alpha^2-p^2|}{p^2}}
        \right) \,.
\end{eqnarray}

\subsection{Numerical results}

We can also solve the integral equations for the wavefunction
renormalization and the vacuum polarization numerically.  Starting
with bare propagators to calculate the vacuum polarization, we solve
Eq.~(\ref{DSeqA}) with this vacuum polarization. Using this solution
for the wavefunction renormalization, we again calculate the vacuum
polarization and repeat this procedure until the solutions for both
$\Pi(p)$ and $A(p)$ converge to a stable solution. In
Fig.~\ref{figsymanal}, we have plotted the analytic solutions for the
wavefunction renormalization as described in the previous section,
together with our numerical result. We see that they agree very well
with each other qualitatively, and that in the ultraviolet region the
$A(p)$ is almost equal to one, as expected, whereas in the infrared
region, it deviates considerably from the perturbative value.  In
particular, it is important to notice that for $p\downarrow 0$, the
wavefunction renormalization vanishes, as also can be seen from
Eq.~(\ref{analres}).

For the other vertex Ans\"atze, we cannot solve the integral equations
analytically. Numerically, we find a very similar behavior for the
wavefunction renormalization for all five different Ans\"atze, as is
shown in Fig.~\ref{figsymfive}(a). The vacuum polarization also seems
to be quite insensitive to the vertex Ansatz, as can be seen from
Fig.~\ref{figsymfive}(b). So it turns out that the exact form of the
Ansatz is not very relevant for the behavior of the fermion
propagator, {\em provided that one uses a consistent approximation
scheme}: consider the coupled equations for the fermion and photon
propagators, and use the same approximation for the full vertex in
both the fermion and the photon DS equation. However, the {\em
deviation} of the vacuum polarization from the perturbative
leading-order behavior
\begin{eqnarray}      
 \Pi(q) &=& \alpha q \,,
\end{eqnarray}
does depend strongly on the Ansatz, see Fig.~\ref{figsymfive}(c). 

It can be explained that the different vertex Ans\"atze lead to the
same behavior for the wavefunction renormalization. In the symmetric
phase, Eqs.~(\ref{DSeqA})--(\ref{DSeqPi}) reduce to
\begin{eqnarray}   \label{DSeqAsym}
 A(p) &=& 1 + \frac{2\,e^2}{p^2} \int\!\frac{{\rm d^3}k}{(2\pi)^3}  
        \frac{f(A(p),A(k),A(q))}{A(k)k^2} 
        \frac{(p\cdot q)(k\cdot q)}{q^2(q^2+\Pi(q))} \,, \\
 \Pi(q) &=& N e^2 \int\!\frac{{\rm d^3}k}{(2\pi)^3} 
   \frac{(2k^2 - 4 k\cdot q - 6(k\cdot q)/q^2) f(A(p),A(k),A(q))}
        {A(k)A(p)\, k^2 p^2 }\,.
\end{eqnarray}
In the ultraviolet region, the propagators, and also the full
vertices, reduce to the bare ones, so the crucial region is the
infrared region. In the (far) infrared region, the vacuum polarization
dominates in the denominator of the kernel of Eq.~(\ref{DSeqAsym}),
$q^2+\Pi(q)$, so we can approximate this denominator by $\Pi(q)$. We
can now observe that there is a cancellation in the infrared region
between the different effects of the vertex Ansatz: the function $f$
enters the integral equation both in the numerator, and, via the
vacuum polarization, in the denominator. The remaining equation for
the wavefunction renormalization resembles the Eqs.~(\ref{apkqsymeq})
and (\ref{prodsymeq}), which we have analyzed in the previous
subsection. This cancellation explains why we get a very similar
behavior for the wavefunction renormalization, using quite different
Ans\"atze for the vertex.

Note that for such a cancellation, the vacuum polarization has to
depend quite strongly on the precise Ansatz, and in
Fig.~\ref{figsymfive}(c) we see that the deviation from the
perturbative behavior is indeed governed by the powers of $A(p)$ in
our vertex Ansatz. Using a bare vertex, the vacuum polarization
becomes quadratic in $1/A$, with a vertex linear in $A$, the vacuum
polarization becomes linear in $1/A$ and with an Ansatz which is
quadratic in the wavefunction renormalization, the vacuum polarization
does not contain any powers of $A$, and is (almost) equal to the
perturbative one.

Finally, we show in Fig.~\ref{figsym24610} the wavefunction
renormalization for several different numbers of fermion flavors.
This shows that there is only a quantitative dependence on the number
of flavors, as long as we stay in the symmetric phase. The other
Ans\"atze yield a similar result.

\subsection{Infrared Behavior}
\label{secirbeh}

As can be seen from Eq.~(\ref{analres}), the wavefunction
renormalization vanishes in the infrared region, in contrast to what
one might expect based on ordinary perturbation theory or the $1/N$
expansion. Such a vanishing behavior could very well be related to an
anomalous dimension for the wavefunction renormalization as suggested
in \cite{AJM90,ApHe81}. It has been argued that the naive perturbative
behavior
\begin{eqnarray}   \label{anpert}
 A_{\hbox{pert}}(p) &\simeq& 1 + \frac{8}{3 N \pi^2} \ln{(p/\alpha)}\,,
\end{eqnarray}
is the first term of the build-up of an anomalous dimension
\begin{eqnarray}   \label{andim}
 A(p) &\simeq& \left(\frac{p}{\alpha}\right)^\eta \,,
\end{eqnarray}
with
\begin{eqnarray}
 \eta &=& \frac{8}{3 N \pi^2} \,.
\end{eqnarray}
An expansion around the origin of the analytic solution
Eq.~(\ref{analres}) gives as leading contribution
\begin{eqnarray} 
  A(p) &=& 1/\left(1 - \frac{8}{3 N \pi^2} \ln{(p/\alpha)}\right) \,.
\end{eqnarray}
So although we do not find the anomalous behavior Eq.~(\ref{andim})
explicitly, our result is in agreement with it to leading order in
$1/N$. Moreover, our solution vanishes at $p \downarrow 0$, just like
the suggested solution Eq.~(\ref{andim}), and does not have the
unphysical behavior of the perturbative result Eq.~(\ref{anpert}),
which diverges at small momenta.

The results with the Ansatz \ref{simple} confirm this anomalous
behavior of the wavefunction renormalization very well. In
Fig.~\ref{figsym24610}, we have also plotted the behavior
Eq.~(\ref{andim}), together with our numerical results using this
Ansatz. This shows that our numerical results are in good agreement
with the expectation of an anomalous dimension for the wavefunction
renormalization, at least in the infrared region. In the ultraviolet
region, Eq.~(\ref{andim}) is not expected to be valid.

%%%%%%%%%%%%%%%%%%%%%%%%%%%%%%%%%%%%%%%%%%%%%%%%%%%%%%%%%%%%%%%%%%%%%%%%%%%%%
\section{Dynamical Mass Generation}
\label{secbroken}

One of the interesting features of QED3 is that a fermion mass can be
generated dynamically, breaking chiral symmetry. Starting with
massless bare fermions, they can acquire a dynamical mass through
nonperturbative effects. The order parameter for this symmetry
breaking is the chiral condensate $\langle\bar\psi\psi\rangle$, but it
is more convenient to consider the infrared value of $B(p)$ as the
order parameter.

Writing the full fermion propagator as
\begin{eqnarray}      
  S^{-1}(p) &=& A(p) {\,\not\!p} + B(p) \,,
\end{eqnarray}
a nonzero solution for $B(p)$ implies a nonzero condensate, and
signals dynamical mass generation. The infrared value of the dynamical
mass function defined by $m(p)=B(p)/A(p)$, $m(0)=B(0)/A(0)$ can also
be used as order parameter. Note we are not calculating the physical
mass, defined at the pole of the propagator; this physical mass is
expected to be in the Minkowski region, at least for observable
particles, whereas we are performing our calculations completely in
Euclidean region. The question of the existence of such a physical
mass in the time-like region is not addressed here \cite{Ma95}.

\subsection{Existence of a critical number of flavors}

Based on bifurcation theory, we can show that, for determining the
critical number (if there is a critical number) of fermion flavors, it
is sufficient to keep the terms which are linear in the generated
mass, assuming that there is a continuous phase transition. This means
that we can use the symmetric solutions for the vacuum polarization
and wavefunction renormalization in the equation for $B$. To avoid
infrared problems, we replace the denominator $A^2(k) k^2 + B^2(k)$ by
$A^2(k) k^2 + B^2(0)$, and thus the equation for $B$ reduces
to\footnote{Alternatively we could neglect the mass function in the
denominator and introduce an infrared cutoff, which is generally
identified with the infrared value of the mass function.}
\begin{eqnarray}  
 B(p) &=& 2\,e^2 \int\!\frac{{\rm d^3}k}{(2\pi)^3} 
        \frac{B(k)}{A^2(k)k^2 + B^2(0)} 
        \frac{f(A(p),A(k),A(q))}{q^2+\Pi(q)} \,,
\end{eqnarray}
or in terms of the mass function
\begin{eqnarray}   \label{DSeqmbif}
 m(p) &=& 2\,e^2 \int\!\frac{{\rm d^3}k}{(2\pi)^3} 
        \frac{m(k)}{k^2 + m^2(0)} 
        \frac{f(A(p),A(k),A(q))}{A(p) A(k) (q^2+\Pi(q))} \,,
\end{eqnarray}
with $A$ and $\Pi$ as found in the previous section.

Consider first the Ans\"atze \ref{apkq} and \ref{prod}. They lead to
\begin{eqnarray}   \label{apkqbroken}
 m(p) &=& 2\,e^2 \int\!\frac{{\rm d^3}k}{(2\pi)^3} 
        \frac{m(k)}{k^2 + m^2(0)} 
        \frac{1}{A(q) q^2+\alpha q} \,,
\end{eqnarray}
and
\begin{eqnarray}    \label{prodbroken}
 m(p) &=& 2\,e^2 \int\!\frac{{\rm d^3}k}{(2\pi)^3} 
        \frac{m(k)}{k^2 + m^2(0)} 
        \frac{1}{q^2+\alpha q } \,,
\end{eqnarray}
respectively. We note that these equations are qualitatively similar:
in the infrared region, the part coming from the vacuum polarization,
$\alpha q$, will dominate over the other part, $A(q) q^2$ or $q^2$
respectively, whereas in the ultraviolet region $A(q) \simeq 1$. Thus,
both Ans\"atze lead to almost the same equation for the mass function. 
Note that Eq.~(\ref{prodbroken}) is exactly the same equation for the
mass function as one would get by using a bare vertex and neglecting
the wavefunction renormalization completely \cite{Appetal}.

Using the fact that the essential region for dynamical mass generation
is the infrared region $p \ll\alpha$, we can use the scale $\alpha$ as
an ultraviolet cutoff. This reduces both Eq.~(\ref{apkqbroken}) and
Eq.~(\ref{prodbroken}) to exactly the same equation, namely
\begin{eqnarray}    \label{appelbroken}
 m(p) &=& \frac{8}{N\,\pi^2} \int_0^\alpha\!\frac{{\rm d}k}{\max(p,k)} 
        \frac{k^2 m(k)}{k^2 + m^2(0)} \,.
\end{eqnarray}
It is well-known that this equation leads to a critical number of
fermion flavors for dynamical mass generation. This is seen directly
by reducing it to a second-order differential equation
\begin{eqnarray}
 p^2 m''(p) + 2 p m'(p) + 
  \frac{8}{N \pi^2} \frac{p^2m(p)}{p^2 + m^2(0)} &=& 0  \,,
\end{eqnarray}
with boundary conditions. Below the critical number of fermion
flavors, $N_c =32/\pi^2 \simeq 3.24$, the solution is given by a
hypergeometric function
\begin{eqnarray}
 m(p) = m(0) \; _2F_1
    \left(a_+,a_-,{\textstyle\frac12}; \frac{p^2}{m^2(0)}\right) \,,
\end{eqnarray}
with 
\begin{eqnarray}
   a_\pm &=& \frac{1}{4} \pm \frac{i}{4}\sqrt{N_c/N-1}\,.
\end{eqnarray}
As $N$ approaches the critical number of fermion flavors, the infrared
value of the mass function $m(0)$ decreases rapidly according to
\begin{eqnarray}    \label{critbeh}
   m(0) &=& \alpha \exp\left[\frac{-2\pi}{\sqrt{N_c/N - 1}}
                + 3\ln2 + {\textstyle\frac12}\pi \right] \,.
\end{eqnarray}
Above the critical number of flavors, the trivial solution $m(p) = 0$
is the only solution.

The equation for the dynamical mass function using the other Ans\"atze
cannot so easily be obtained. However, using a simple counting
argument, we can already expect that there is a critical number of
fermion flavors, independent of the precise form of our Ansatz. In the
infrared region, which is essential for dynamical symmetry breaking,
we can neglect $q^2$ with respect to the vacuum polarization in the
denominator $q^2 + \Pi(q)$. Counting the powers of $A$, we see that
this vacuum polarization is roughly proportional to $f/A^2$, so the
dependence of the integration kernel in Eq.~(\ref{DSeqmbif}) on the
wavefunction renormalization and the vertex Ansatz might cancel out. 
This is similar to what we have already seen in the previous section,
namely that the behavior of the wavefunction renormalization is almost
independent of the vertex Ansatz we use.

Therefore, although the behavior of the vacuum polarization might
depend on the vertex Ansatz, we expect that in fact the fermion
propagator is not very sensitive to the precise Ansatz, due to a
cancellation in the (far) infrared region between the implicit
dependence of $\Pi(q)$ on the function $f(A(p),A(k),A(q))$, and the
explicit appearance of this function in the integration kernel of
Eq.~(\ref{DSeqmbif}). On these grounds one might expect a critical
number of fermion flavors, independent of our choice for the vertex.

Even if we relax our requirement that the Ansatz goes to the bare
vertex in the ultraviolet region, we find a similar result: consider
for example the (unphysical) vertex
\begin{eqnarray}
   2\,A(p)A(k) \gamma^\mu \,.
\end{eqnarray}
This would lead to the equation for the mass function like
\begin{eqnarray}
 m(p) &=& 2\,e^2 \int\!\frac{{\rm d^3}k}{(2\pi)^3} 
        \frac{m(k)}{k^2 + m^2(0)} 
        \frac{2}{q^2+ 2\alpha q} \,,
\end{eqnarray}
which has an additional factor of two. However, in the infrared
region, this factor of two cancels against the additional factor of
two which comes from the vacuum polarization, leading to the same
critical number as with Ansatz \ref{prod}.

Note that it is crucial for such a cancellation to occur that we
consider the vacuum polarization with full propagators; using bare
propagators in the loop for the vacuum polarization, is an
inconsistent approximation. Formally, using such bare fermions in the
loop for the vacuum polarization is in agreement with the $1/N$
expansion, but as we have seen in the previous section, the
wavefunction renormalization is not of the order $1+{\cal O}(1/N)$ in
the infrared region. Therefore we should use the full propagators when
calculating the vacuum polarization. Even when using bifurcation
theory to calculate the critical number of fermion flavors, one should
use the full propagators in the vacuum polarization, or to be more
precise, the massless full propagators.

\subsection{Numerical results}

We can solve the coupled equations numerically for the mass function
$m$ (or for $B$), the wavefunction renormalization $A$, and the vacuum
polarization $\Pi$ in the broken phase. Starting with a trial function
for $B$ and the leading order contribution for the wavefunction,
$A(p)=1$, we can evaluate the vacuum polarization and solve the
coupled equations for $A$ and $B$. Next, we calculate the vacuum
polarization, using these numerical solutions, and iterate this
procedure until all three functions converge to a stable solution.

For the quenched approximation, $N = 0$, all five Ans\"atze lead to
the same solution for $B$, since the vacuum polarization is zero in
this case, and the wavefunction renormalization equal to one. This
mass function is almost constant in the infrared region, and decreases
rapidly, as $1/p^2$, at large momenta, for $p > e^2$. It agrees very
well with earlier numerical studies of this approximation
\cite{HoMa89}.

The typical behavior of the functions $A$, $B$, and $\Pi$ is shown in
Fig.~\ref{figABPibroken} for several values of $N$. We note that the
wavefunction renormalization decreases for small momenta, but does not
vanish for $p\downarrow 0$: its value at the origin is nonzero, in
contrast to the behavior in the symmetric phase. Also the behavior of
the vacuum polarization is different: in the broken phase, the vacuum
polarization behaves like $p^2$ in the (far) infrared region, so
$\Pi(p)/p^2$ is finite at $p\downarrow 0$, whereas in the symmetric
phase, $\Pi(p)/p^2$ blows up at the origin. The infrared behavior of
the vacuum polarization is governed by the generated mass function:
the smaller this mass function, the larger $\Pi(p)/p^2$. Finally the
mass function, or rather $B$, is almost constant at small momenta, and
decreases rapidly for large momenta. However, in the unquenched
approximation, and especially close to the critical number of fermion
flavors, we see {\em two} relevant mass scales \cite{GuHaRe95}: $B(p)$
starts to decrease at the scale of the generated mass, but only beyond
the energy scale $\alpha$, it decreases like $1/p^2$. This phenomenon
might be relevant for hierarchy problems in unified theories.

To determine the critical number of fermion flavors, we consider $N$
as a continuous parameter in the DS equations. In
Fig.~\ref{figABPivsN}, we show the infrared values of the wavefunction
renormalization, of the scalar function $B$, and of the vacuum
polarization, or rather of $\Pi(p)/p^2$, as functions of $N$. For
increasing $N$, we note that $B(0)$ decreases, and that the infrared
value of $\Pi(p)/p^2$ increases rapidly. Both functions indicate a
critical value of the number of fermion flavors for dynamical mass
generation, at which $B$ vanishes, and where the limit $p\downarrow 0$
of $\Pi(p)/p^2$ diverges. Also the wavefunction renormalization at the
origin decreases quite rapidly close to this critical number of
flavors.

The dependence of $m(0)$ as function of $N$ at fixed $\alpha$ is shown
in Fig.~\ref{figcritical}. For comparison, we also included the
analytical result Eq.~(\ref{critbeh}). We see very clearly that the
generated infrared mass also decreases very rapidly for increasing
$N$, and the figure indicates a critical number of fermion flavors of
about $3.3$, more or less independent of the vertex Ansatz. The
Ans\"atze \ref{apkq}, \ref{quad}, and \ref{prod} give exactly the same
critical behavior, and also Ansatz \ref{simple} leads to almost the
same results. Only a bare vertex, Ansatz \ref{bare}, indicates a
slightly higher value for the critical number of flavors, but the
general behavior is the same for all five Ans\"atze.

%%%%%%%%%%%%%%%%%%%%%%%%%%%%%%%%%%%%%%%%%%%%%%%%%%%%%%%%%%%%%%%%%%%%%%%%%%%%%
\section{Conclusion}
\label{secconc}

We have solved the {\em coupled} DS equations for the fermion
propagator and the vacuum polarization, both in the chirally symmetric
and in the broken phase, using a certain class of vertex functions. 
With this type of vertex, the behavior of the fermion propagator is
almost independent of the exact form of the full vertex.  We find a
critical number for chiral symmetry breaking, $N_c = 3.3$, below which
there is dynamical mass generation; above this critical number, only
the chirally symmetric solution exists.

In the chirally symmetric phase, the wavefunction renormalization is
approximately equal to one in the ultraviolet region, but in the
infrared region, $A(p)$ vanishes, indicating an anomalous dimension. 
This non-uniform behavior of the wavefunction renormalization lies at
the origin of a controversy about the existence of a critical number
of fermion flavors. Also in the broken phase, $A(p)$ is not of the
order one, as one might expect naively, but considerably smaller: as
the number of fermion flavors approaches the critical number, $A(0)$
tends to go to zero. These results will also be relevant for studies
of this model at finite temperature \cite{AiKl94}.

Our main result, a finite critical number of fermion flavors for
dynamical mass generation, confirms the assertions obtained earlier by
Appelquist {\em et al.} \cite{Appetal}. Their analysis was based upon
a bare vertex and neglect of the wavefunction renormalization. There
have been several claims that including the effects of the
wavefunction renormalization leads to a different result, namely
dynamical mass generation for all numbers of fermion flavors
\cite{Penetal88,PW91CPW92}. These and other studies, such as
\cite{AJM90,KoNa92}, show a crucial dependence on the behavior of the
full vertex; in general a bare vertex or a vertex linear in the
wavefunction renormalization would lead to chiral symmetry breaking
for all $N$, whereas a vertex quadratic in $A$ leads to a critical
number of fermion flavors.  However, all these studies include the
wavefunction renormalization and an Ansatz for the full vertex in the
fermion DS equation, but not in the vacuum polarization.

We would like to stress that in order to solve the DS equation for the
fermion propagator self-consistently, we have to treat the fermion
propagator nonperturbatively {\em both} in the fermion DS equation
{\em and} in the equation for the vacuum polarization, at least if we
are using the unquenched approximation. Also, one must use the same
approximation for the full vertex in both the fermion and photon DS
equation. It turns out that the effects of the wavefunction
renormalization through the vacuum polarization change drastically the
naive results obtained in \cite{Penetal88,PW91CPW92,AJM90,KoNa92}. In
contrast to these earlier results, our results are almost independent
of the vertex Ansatz.

For studying the chiral phase transition, one can use bifurcation
theory and therefore neglect the dynamical mass function in the
wavefunction renormalization. However, one should keep in mind that
bifurcation theory does not imply that one can use bare propagators in
the vacuum polarization: one should use the chirally-symmetric full
propagators. Of course, if the wavefunction renormalization is equal
to one (as one often assumes in this kind of calculations), one can
use the bare propagators to calculate the vacuum polarization.

We have made only one approximation in our calculation: replacing the
full vertex by our Ansatz, Eq.~(\ref{genansatz}). Of course, one could
question this approximation, but our result is almost independent of
the precise form of the function $f$. Furthermore, it has been shown
that a next-to-leading order calculation in the context of the $1/N$
expansion also leads to a finite critical number of fermion flavors
\cite{Na89}. Our result is also in good agreement with lattice
calculations \cite{DKK89}.

%%%%%%%%%%%%%%%%%%%%%%%%%%%%%%%%%%%%%%%%%%%%%%%%%%%%%%%%%%%%%%%%%%%%%%%%%%%%%
\section*{Acknowledgements}

I would like to thank Yoonbai Kim, Conrad Burden, K.~Yamawaki, and
Y.~Hoshino for stimulating discussions and comments. This work has
been financially supported by the Japanese Society for the Promotion
of Science (JSPS fellowship No. 94146) and by a Grand-in-Aid for
Scientific Research from the Japanese Ministry of Education, Science,
and Culture (No. 90094146).

%%%%%%%%%%%%%%%%%%%%%%%%%%%%%%%%%%%%%%%%%%%%%%%%%%%%%%%%%%%%%%%%%%%%%%%%%%%%%
\begin{figure}
\caption{The the numerical and analytical solutions for the 
         wavefunction renormalization in the massless phase for $N=6$,
         using the Ans\"atze \protect{\ref{apkq}} and \protect{\ref{prod}}.}
\label{figsymanal}
\vspace{1cm}
\caption{The numerical solutions in the massless phase for $N=6$,
         using the Ans\"atze \protect{\ref{bare}}-\protect{\ref{prod}}:
         (a) the wavefunction renormalization, 
         (b) the vacuum polarization, 
         and (c) the deviation of the vacuum polarization 
         from the perturbative behavior.} 
\label{figsymfive}
\vspace{1cm}
\caption{The wavefunction renormalization in the massless phase 
         for $N = 2$, $4$, $6$, and $10$, using Ansatz
         \protect{\ref{simple}}, together with the behavior based 
         on the existence of an anomalous dimension.}
\label{figsym24610}
\vspace{1cm}
\caption{The functions $A(p)$, $B(p)$, and $\Pi(p)/p^2$ 
        in the chirally broken phase for N = 0, $2$, and $3$, 
        using Ansatz \protect{\ref{simple}}.}
\label{figABPibroken}
\vspace{1cm}
\caption{The infrared values of the functions $A(0)$, $B(0)$, 
        and $\lim_{p\rightarrow 0} \Pi(p)/p^2$
        as functions of $N$ at fixed $e^2$}
\label{figABPivsN}
\vspace{1cm}
\caption{The infrared values of the mass function $m(0)$ as function 
         of $N$ at fixed $\alpha$.}
\label{figcritical}
\end{figure}

%%%%%%%%%%%%%%%%%%%%%%%%%%%%%%%%%%%%%%%%%%%%%%%%%%%%%%%%%%%%%%%%%%%%%%%%%%%%%


\begin{thebibliography}{99}
\bibitem[*]{emmar} 
 e-mail: maris@eken.phys.nagoya-u.ac.jp

\bibitem{Pi84} 
  R.D. Pisarski, 
  Phys. Rev. D {\bf 29}, 2423 (1984).

\bibitem{Appetal}
  T. Appelquist, M. Bowick, D. Karabali, and L.C.R. Wijewardhana, 
  Phys. Rev. D {\bf 33}, 3704 (1986);
  T. Appelquist, D. Nash, and L.C.R. Wijewardhana, 
  Phys. Rev. Lett. {\bf 60}, 2575 (1988).

\bibitem{Na89}
  D. Nash, 
  Phys. Rev. Lett. {\bf 62}, 3024 (1989).

\bibitem{HoMa89}
  Y. Hoshino and T. Matsuyama, 
  Phys. Lett. B {\bf 222}, 493 (1989).

\bibitem{DKK89}
  E. Dagotto, J.B. Kogut, and A. Koci\'c, 
  Phys. Rev. Lett. {\bf 62}, 1083 (1989); 
  E. Dagotto, A. Koci\'c, and J.B. Kogut, 
  Nucl. Phys. B {\bf 334}, 279 (1990).

\bibitem{Penetal88}
  M.R. Pennington and S.P. Webb, 
  {\it Hierarchy of scales in three dimensional QED},
  BNL-40886, January 1988 (unpublished);
  D. Atkinson, P.W. Johnson and M.R. Pennington, 
  {\it Dynamical mass generation in three-dimensional QED},
  BNL-41615, August 1988 (unpublished).

\bibitem{PW91CPW92}
  M.R. Pennington and D. Walsh, 
  Phys. Lett. B {\bf 253}, 246 (1991);
  D.C. Curtis, M.R. Pennington, and D. Walsh, 
  Phys. Lett. B {\bf 295}, 313 (1992).

\bibitem{AJM90}
  D. Atkinson, P.W. Johnson, and P. Maris, 
  Phys. Rev. D {\bf 42}, 602 (1990).

\bibitem{KoNa92}
  K.-I. Kondo and H. Nakatani,
  Progr. Theor. Phys. {\bf 87}, 193 (1992).

\bibitem{BuRo91}
  C.J. Burden and C.D. Robert,
  Phys. Rev. D {\bf 44}, 540 (1991).

\bibitem{BuPrRo92}
  C.J. Burden, J. Praschifka, and C.D. Roberts,
  Phys. Rev. D {\bf 46}, 2695 (1992).

\bibitem{GrSeSo95}
  G. Grignani, G. Semenoff, and P. Sodano,
%%  {\em Confinement-deconfinement transition in three-dimensional QED},
  DFUPG-100-95, hep-th/9504105.

\bibitem{Ma95}
  P. Maris, hep-ph/9508323,
  Phys. Rev. D {\bf 52}, 6087 (1995).  

\bibitem{DoMa}
  N. Dorey and N.E. Mavromatos,
  Nucl. Phys. B {\bf 386}, 614 (1992);
  I.J.R. Aitchison and N.E. Mavromatos, OUTP-95-32p, hep-th/9510058.

\bibitem{AbCa95}
  E. Abdalla and F.M. de Carvalho Filho,
  MIT-CTP-2488, IC/95/350, hep-th/9511132.

\bibitem{ABKW86}
  T. Appelquist, M.J. Bowick, D. Karabali, and L.C.R. Wijewardhana, 
  Phys. Rev. D {\bf 33}, 3774 (1986).

\bibitem{Miretal}
  P.I. Fomin, V.P. Gusynin, V.A. Miransky and Yu.A. Sitenko,
  Nuovo Cimento {\bf 6}, 1 (1983);
  V.A. Miransky, Phys. Lett. B {\bf 146}, 401 (1985);
  Nuovo Cimento {\bf 90A}, 149 (1985).

\bibitem{Si90}
  E.H. Simmons, 
  Phys. Rev. D {\bf 42}, 2933 (1990). 

\bibitem{KM94}
  K.-I. Kondo and P. Maris,
  hep-ph/9408210, Phys. Rev. Lett. {\bf 74}, 18 (1995);
  hep-ph/9501280, Phys. Rev. D {\bf 52}, 1212 (1995).

\bibitem{GuHaRe95}
  V.P. Gusynin, A.H. Hams, and M. Reenders,
%%  {\em (2+1)-dimensional QED with dynamically massive fermions 
%%   in the vacuum polarization}, 
  hep-ph/9509380, Phys. Rev. D {\bf 53}, 2227 (1996).

\bibitem{Ko95}
  K.-I. Kondo, 
  hep-ph/9509345, Int. J. Mod. Phys. A {\bf 11}, 777 (1996).  

\bibitem{epsilon}
  R.D. Pisarski, Phys. Rev. D {\bf 44}, 1866 (1991); 
  G.W. Semenoff, P. Suranyi, and L.C.R. Wijewardhana, hep-ph/9502272. 

\bibitem{KiKi96}
  Seyong Kim and Yoonbai Kim,
  {\em Lattice Gauge Theory of Three-dimensional Thirring Model},
  SNUTP-96-010.

\bibitem{BaCh80}
  J.S. Ball and T.W. Chiu,
  Phys. Rev D {\bf 22}, 2542 (1980).

\bibitem{KiRePe95}
  A. Kizilers\"u, M. Reenders, and M.R. Pennington,
  hep-ph/9503238, Phys. Rev. D {\bf 52}, 1242 (1995).

\bibitem{ApHe81}
  T.W. Appelquist and U. Heinz, 
  Phys. Rev. D {\bf 24}, 2169 (1981).

\bibitem{AiKl94}
  I.J.R. Aitchison and M. Klein-Kreisler,
  hep-ph/9402213, Phys. Rev. D {\bf 50}, 1068 (1994).

\end{thebibliography}
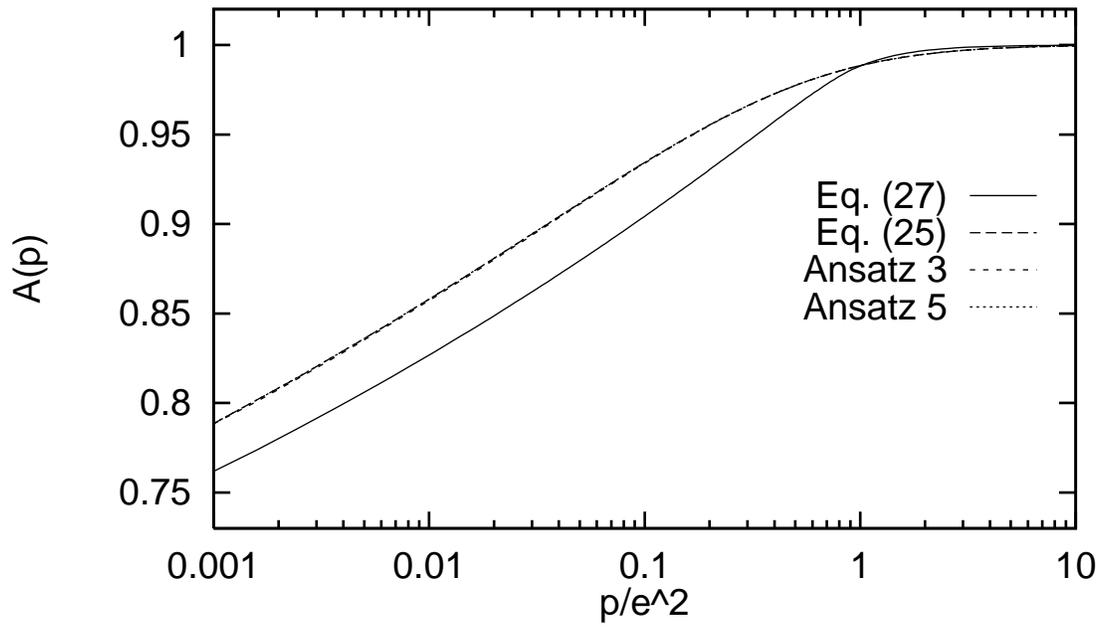
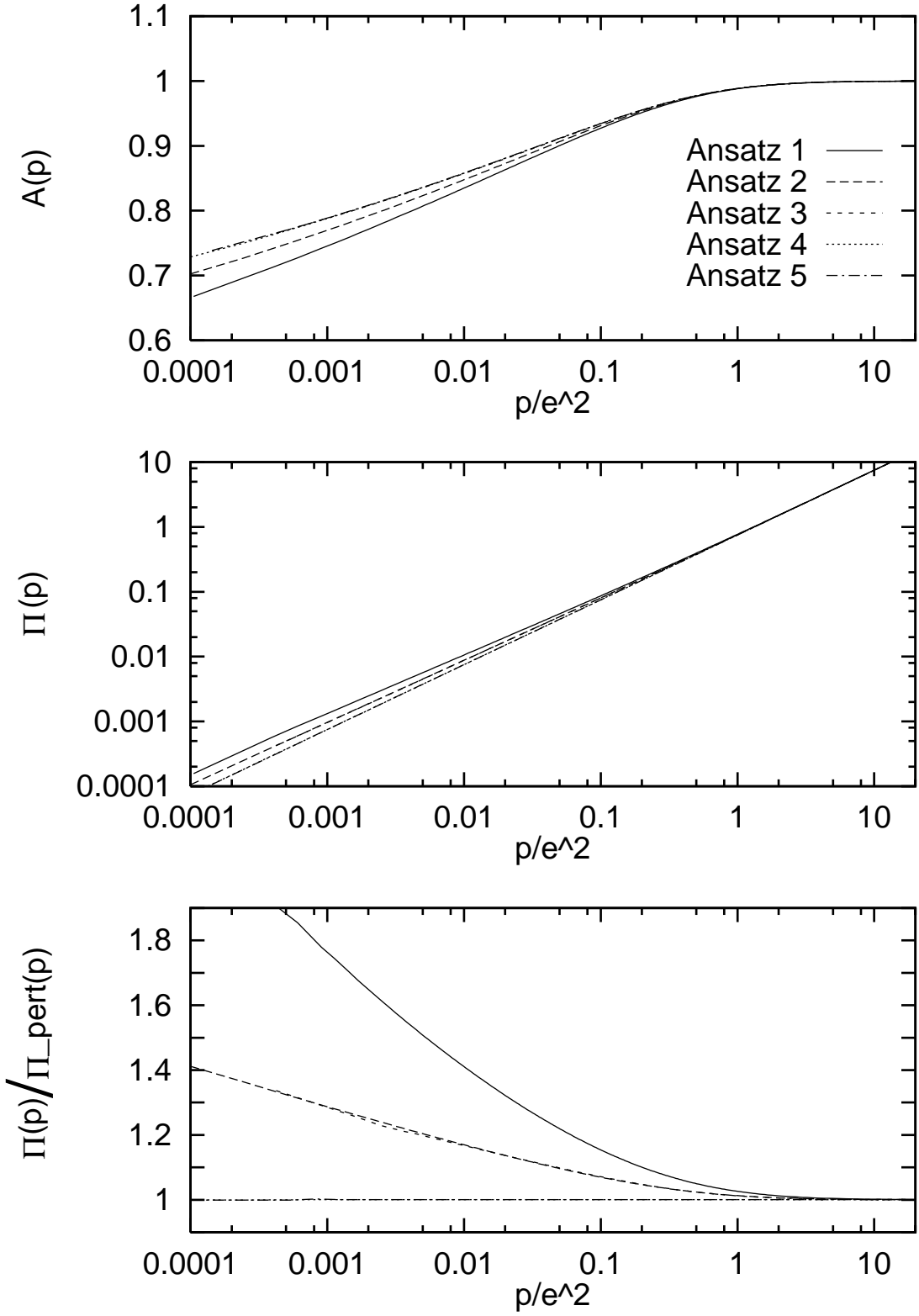
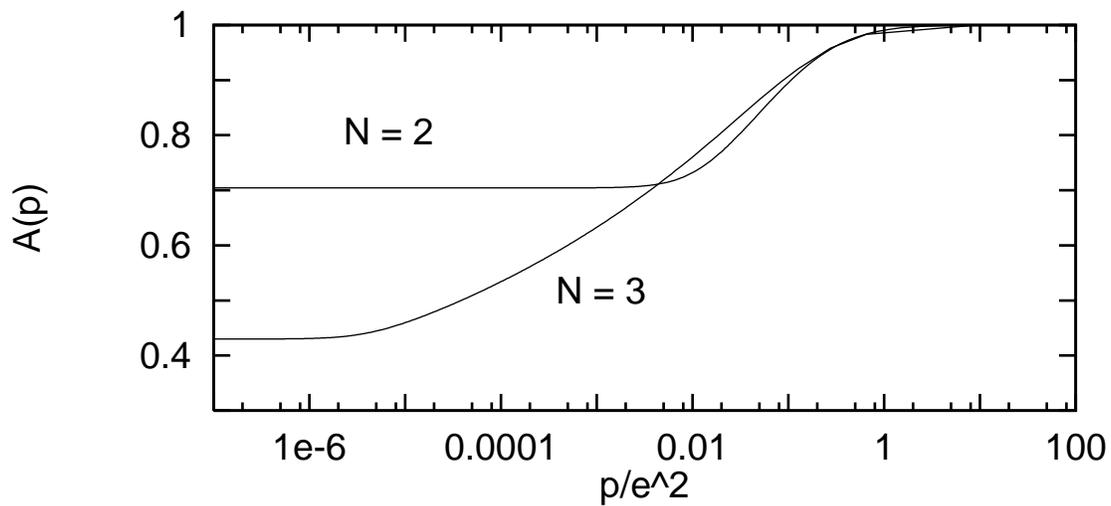
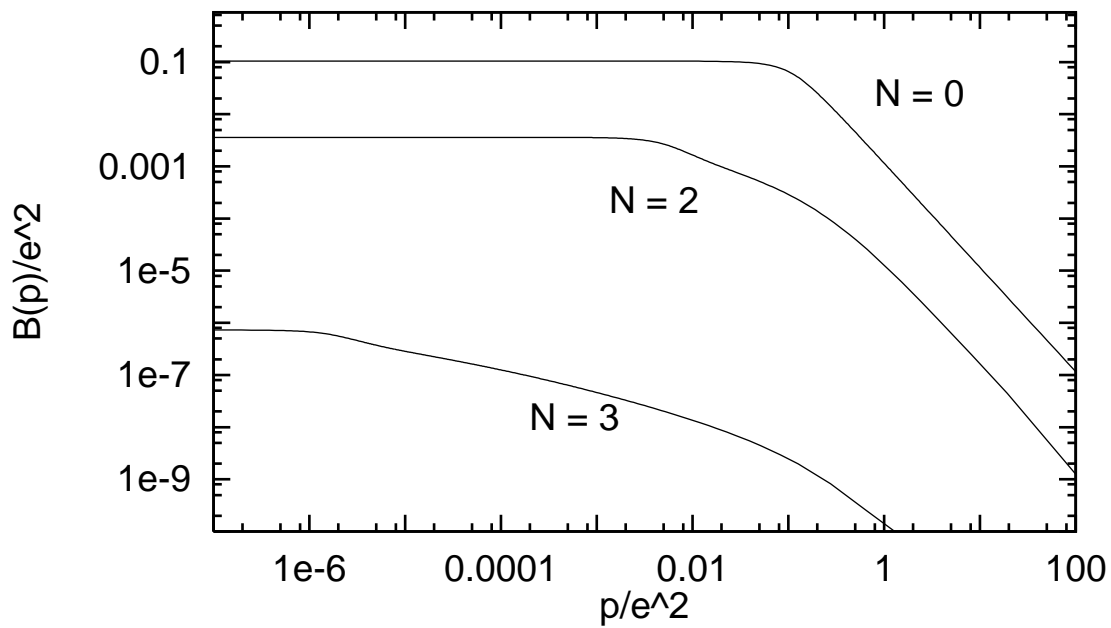
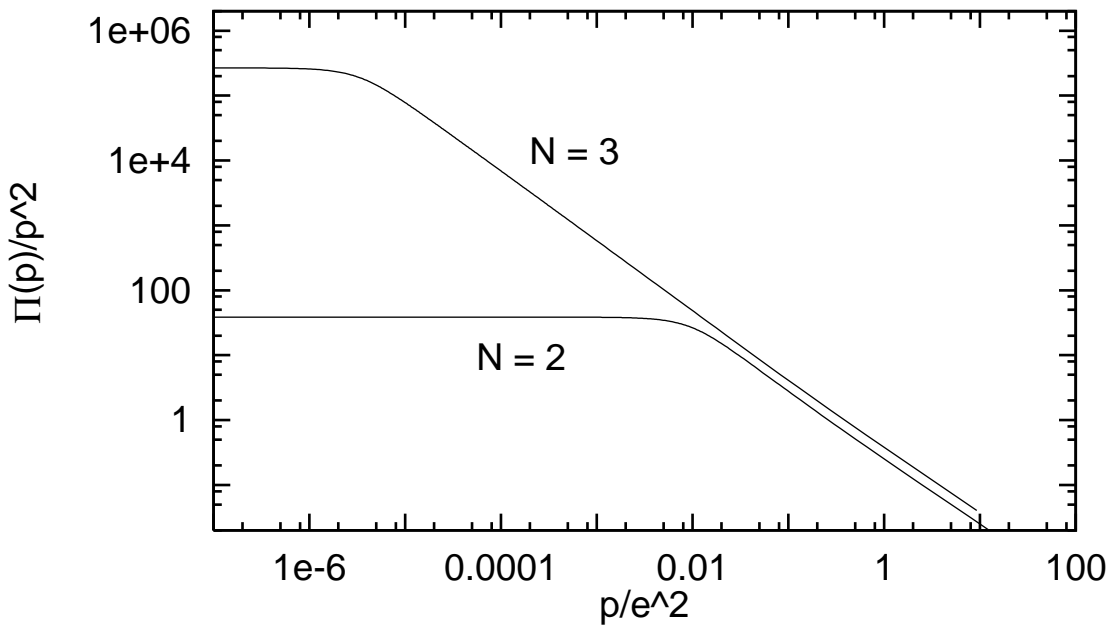
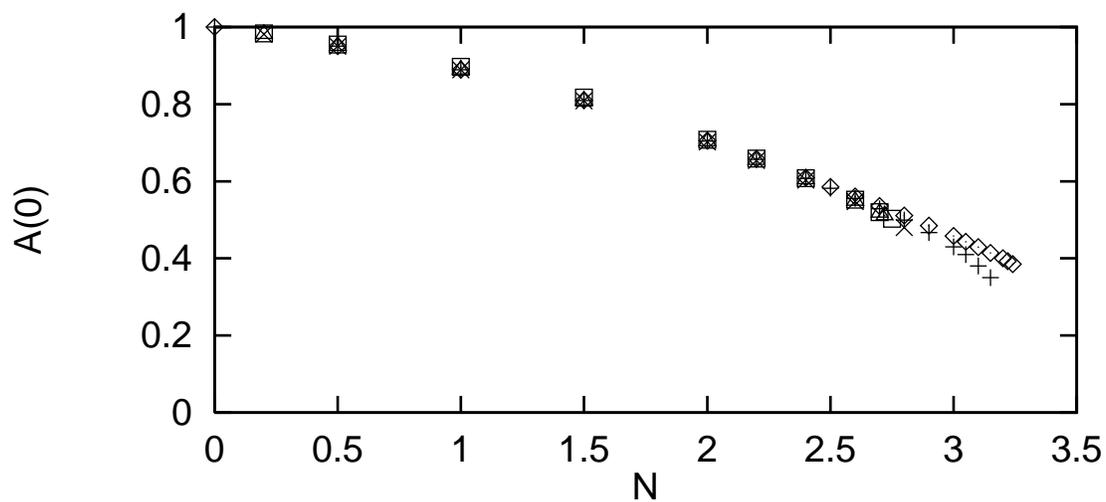
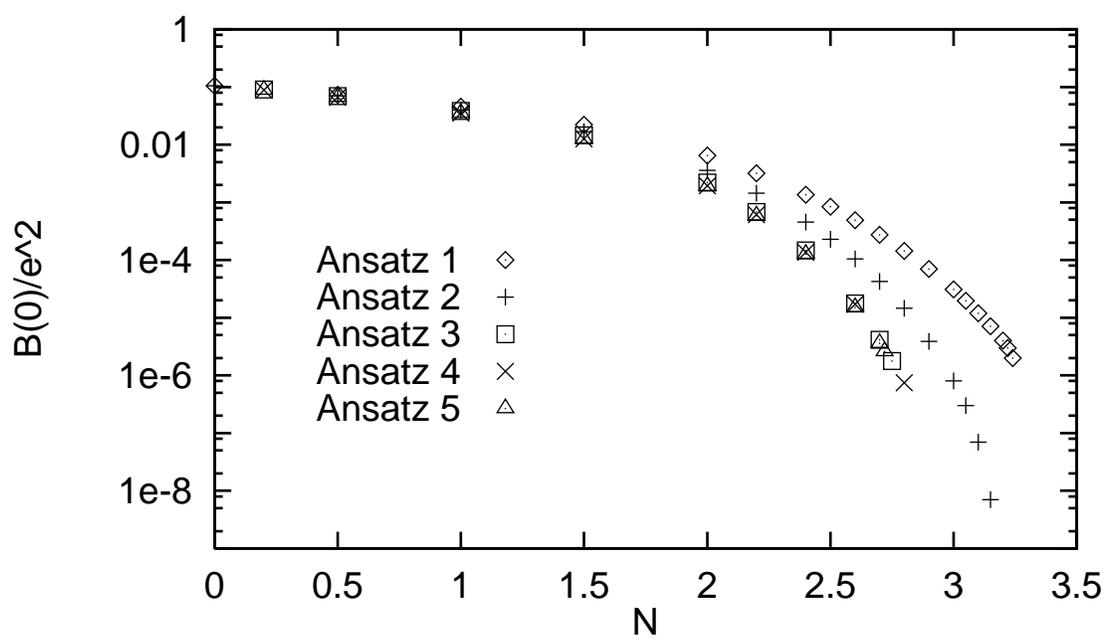
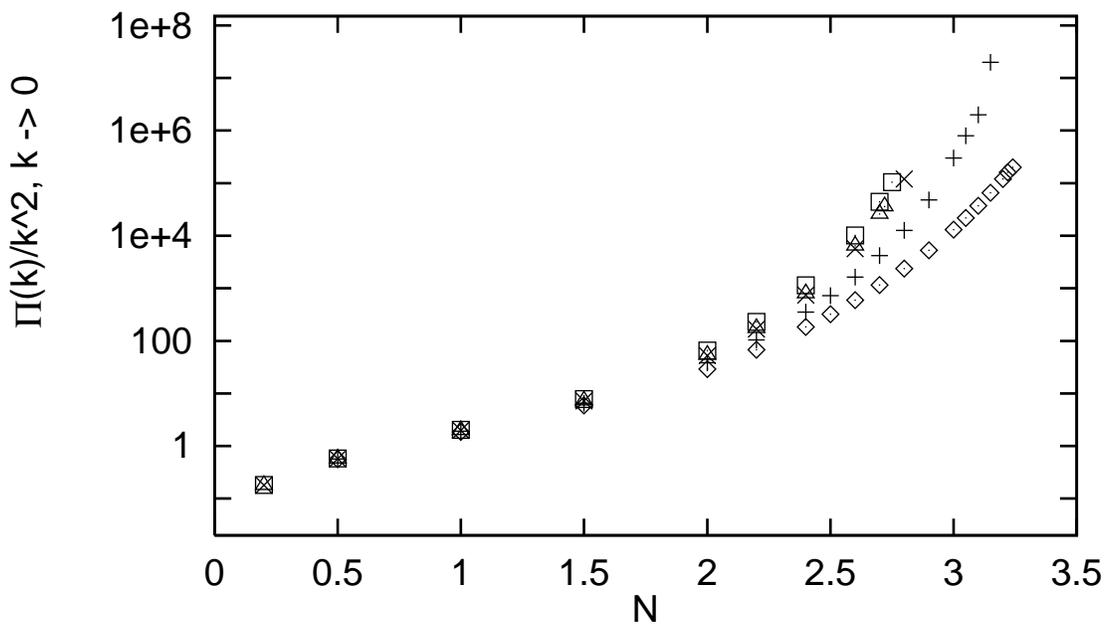
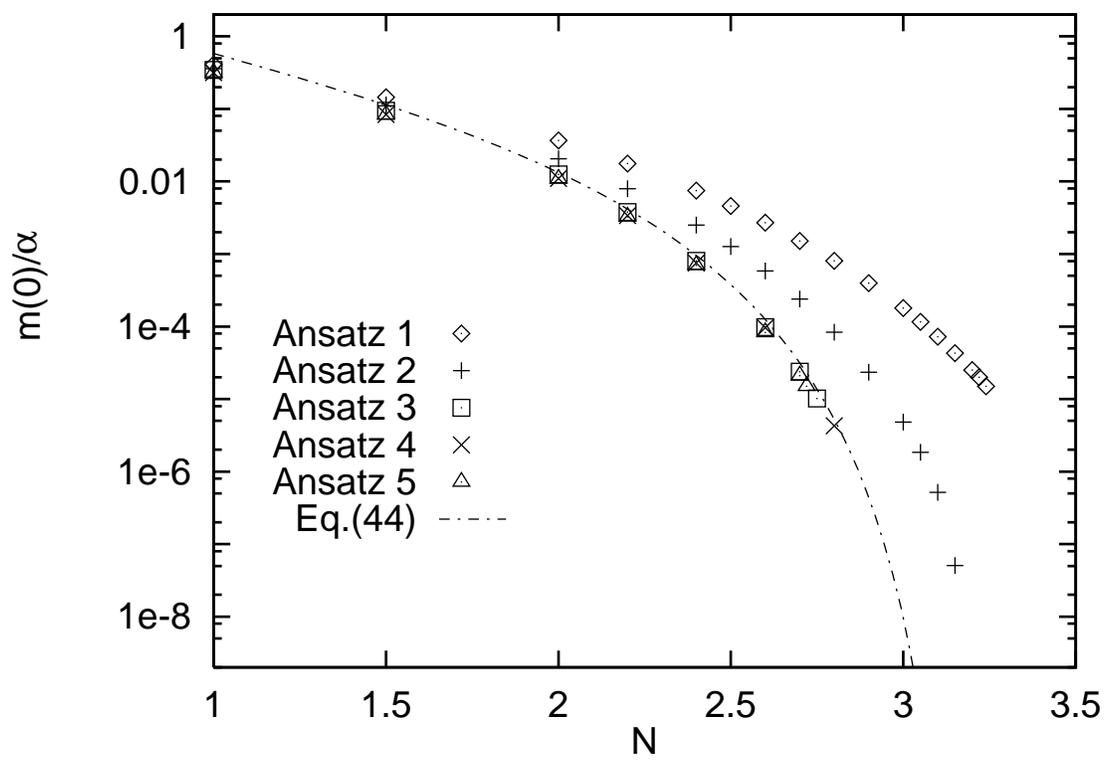
\end{document}